# Erbium-doped lithium niobate thin film waveguide amplifier with 16 dB internal net gain

Minglu Cai, Kan Wu,* Junmin Xiang, Zeyu Xiao, Tieyin Li, Chao Li, and Jianping Chen

**ABSTRACT:** Erbium-doped lithium niobate on insulator (Er:LNOI) has attracted enormous interest as it provides gain and enables integrated amplifiers and lasers on the lithium niobate on insulator (LNOI) platform. We demonstrate a highly efficient waveguide amplifier on Er:LNOI. The 2.58-cm long amplifier can achieve 27.94 dB signal enhancement, 16.0 dB internal net gain (6.20 dB/cm), -8.84 dBm saturation power, 4.59 dB/mW power conversion efficiency, and 4.49 dB noise figure at 1531.6 nm. Besides, thorough investigation on the pumping wavelength, pumping scheme, output power and noise figure have been performed to provide a comprehensive understanding on this novel waveguide amplifier. This work will benefit the development of a powerful gain platform and can pave the way for a fully integrated photonic system on LNOI platform.

## 1. INTRODUCTION

Erbium-doped fiber amplifier has been a great success for the past decades as it provides high-gain and low-noise amplification for the optical signal, which forms the cornerstone for current worldwide optical fiber communication network. Compared to the integrated semiconductor optical amplifiers, the long upper-state lifetime of erbium ions also enables a low-noise amplifier on the chip, which triggers the extensive investigation on the erbium-doped waveguide amplifiers. Various host materials have been chosen including lithium niobate (LN), phosphate, silicate, aluminum oxide and tellurium oxide [1-5], etc. Good performance has been reported, e.g., a maximum internal net gain of 27 dB from an 8.6-cm-long erbium-doped phosphate waveguide [4], and the internal net gain per unit length greater than 100 dB/cm from a 56-μm-long single-crystal erbium chloride silicate nanowire [5]. However, it is still challenging for these erbium platforms to simultaneously achieve high internal net gain and high gain per unit length, which are important for a compact and useful integrated optical amplifier.

Recently, erbium doped lithium niobate on insulator (Er:LNOI) has attracted intense interest because it can simultaneously achieve high erbium doping concentration, good mode confinement and low-loss waveguide. Therefore, integrated optical amplifiers based on Er:LNOI platform can potentially support high total internal net gain and high internal net gain per unit length. Moreover, lithium niobate on insulator (LNOI) has proved to be a powerful platform for various photonic applications [6, 7] including electro-optic modulation, microcomb, supercontinuum generation, second-harmonic generation, optical parametric oscillation, acousto-optic devices [8-14], etc. By adding gain property to the LNOI platform, many new devices and systems are empowered such as a fully integrated photonic system with light sources, modulators, amplifiers, tunable filters and many other active or passive components on a single LNOI chip. As a result, considerable research works have emerged on the theme of erbium-doped LNOI lasers and amplifiers. For Er:LNOI lasers, Liu, Wang and Yin have respectively reported pioneer works on micro-disk and micro-ring with multi-wavelength output [15-17]. Zhang, Gao, Xiao and Lin adopted coupled cavities for single-wavelength lasing [18-21]. Li has later demonstrated that single-wavelength lasing can also be obtained on a single resonator by employing mode-dependent loss and gain filtering [22]. For Er:LNOI waveguide amplifiers, Chen has achieved 5.2 dB internal net gain [23]. Yan has reported 8.3 dB internal net gain and Luo has reported 15 dB internal net gain [24, 25]. Very recently, Zhou has achieved an amplifier with 18 dB internal net gain and 5 dB/cm gain per unit length on a 3.6-cm-long Er:LNOI waveguide [26]. Although many works of Er:LNOI amplifiers have emerged, a comprehensive picture of an Er:LNOI amplifier has not be established yet. That is, for an optical amplifier, other important parameters should also be characterized such as effect of pumping wavelength and pumping scheme, output power and noise figure. Only with the knowledge of all these parameters, an optical amplifier can be comprehensively evaluated.

In the work, we have demonstrated an Er:LNOI waveguide amplifier with a signal enhancement of 27.94 dB, an internal net gain of 6.20 dB/cm (16.0 dB total), a saturation power of -8.84 dBm and a noise figure of 4.49 dB at 1531.6 nm. Moreover, many important parameters have also been investigated including the effect of pumping wavelength and pumping scheme, output power, power conversion efficiency, etc. With this work, a comprehensive understanding of a high-performance Er:LNOI waveguide amplifier can be established, which clearly paves the way for a fully integrated photonic system on LNOI platform.

## 2. DEVICE FABRICATION AND CHARACTERIZATION

The main fabrication procedures of the monolithically integrated waveguide amplifier on Er:LNOI are briefly explained as follows. In Fig. 1(a), the high-quality Z-cut Er:LNOI wafer was prepared by “ion-cutting” technology (NANOLN). The wafer consists of 600 nm-thick erbium-doped lithium niobate (LN) thin film, 2 μm-thick $SiO_2$, and 400 μm-thick Si substrate. The purity property and element proportion of the Er:LN thin film were characterized by X-ray photoelectron spectroscopy (XPS), as shown in Fig. 1(b). The mainly concerned core lines are O 1s, Nb 3d, Li 1s, and Er 4d. In the binding energy range of 170-175 eV, the peak position corresponds to Er 4d, and it can be regarded as the third oxidation state of erbium [27]. The $Er^{3+}$ concentration of LN thin film is calculated as $0.72\times10^{20}$ $cm^{-3}$.

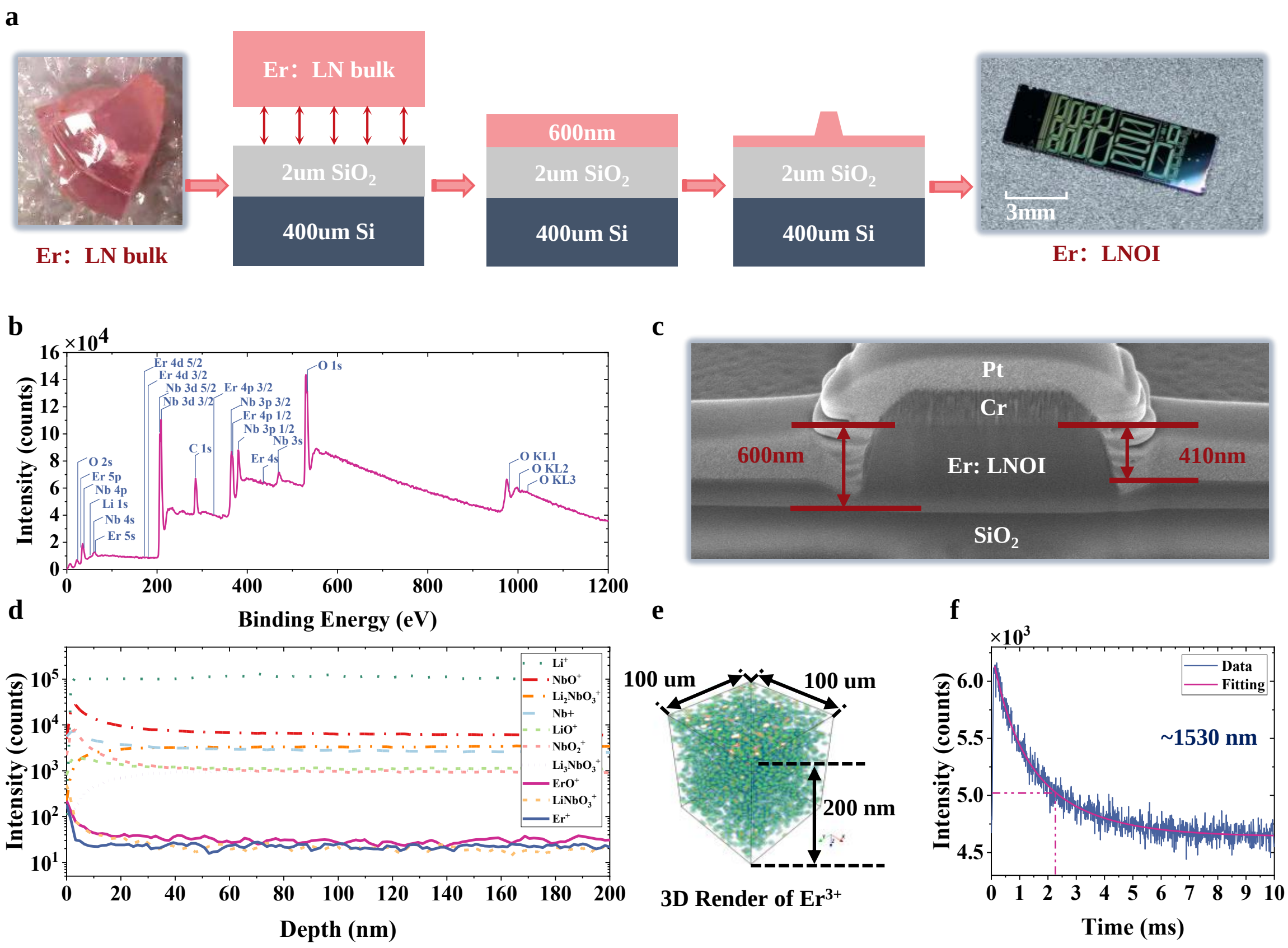

**Figure 1.** (a) Illustration of fabrication of Er:LNOI waveguide amplifiers. (b) The material elements measured by XPS. (c) SEM image of waveguide end facet with Cr mask and Pt. (d) Element distribution measured by TOF-SIMS. (e) $Er^{3+}$ distribution. (f) Fluorescence lifetime

Then a layer of chromium (Cr) was deposited on the surface of Er:LNOI and a layer of hydrogen-silsesquioxane (HSQ) photoresist were spin-coated on the Cr thin film. Through the electron beam lithography (EBL) technology and standard dry etching process, the Cr hard mask was achieved. Next the Er:LNOI ridge waveguides were fabricated by inductively coupled plasma-reactive ion etching (ICP-RIE) technology and the etching gas was argon. The total waveguide height is 600 nm with a ridge height of 410 nm and a slab height of 190 nm. The waveguide end facets were polished by focused ion beam (FIB). Finally, the Cr film was removed to obtain clean Er:LNOI waveguides. Spiral waveguide amplifiers with different lengths were fabricated for comparison. The scanning electron microscopic (SEM) image of the FIB polished waveguide end facet is shown in Fig. 1(c). The coupling loss is approximately 7.2 dB/facet near 1530 nm.

Before discussing the properties of our Er:LNOI amplifiers, we have characterized material properties of Er:LNOI wafer. The element distribution of the Er:LN thin film was characterized by ion time-of-flight secondary ion mass spectrometer (TOF-SIMS), as shown in Fig. 1(d). Within the depth of 200 nm, various element ions uniformly distribute along the depth, which lays the foundation for the high-performance waveguide amplifier. Three-dimensional (3D) distribution of $Er^{3+}$ ions is shown in Fig. 1(e). For the 1530 nm emission ($^4I_{13/2} \rightarrow {}^4I_{15/2}$), the measured fluorescence lifetime $\tau_{pl}$ of $Er^{3+}$ is ~2.3 ms, which is the time corresponding to the attenuated fluorescence intensity of 1/e, as shown in Fig. 1(f).

## 3. EXPERIMENTAL RESULES

### A. Signal Enhancement and Internal Net Gain

The experimental setup is shown in Fig. 2. The signal light is from a C-band tunable laser and the pump light is from a fiber-pigtailed 980 nm laser diode or a homemade 1484 nm Raman fiber laser. Three pumping schemes are to be investigated, i.e., forward, backward, and bi-directional pumping. For forward pumping, signal light and pump light are multiplexed by a wavelength division multiplexer (WDM), propagate through a polarization controller (PC), and couple to the Er:LNOI chip via a lensed fiber. The output of integrated Er:LNOI amplifier is collected by the other lensed fiber. The amplified signal and residual pump are separated by another WDM. The filtered signal is fed into an optical spectrum analyzer (OSA) for spectral measurement. For backward pumping, the WDM in the output side is used to combine the signal and pump, and the WDM in the input side is used to separate them. For bi-directional pumping, the pump is split by a 50:50 fiber coupler and simultaneously enters the device through two WDMs. Micrograph of the fabricated chip is shown in Fig. 2(a). Fig. 2(b) shows a photo of the Er:LNOI amplifier when the pump light is applied. The spiral waveguide is lightened up by the green fluorescence.

The highest gain was achieved in a device with a 2.58-cm-long waveguide and a signal wavelength of 1531.6 nm by bi-directional pumping at 1484 nm. The measured signal enhancement and calculated internal net gain are shown in Fig. 3(a).

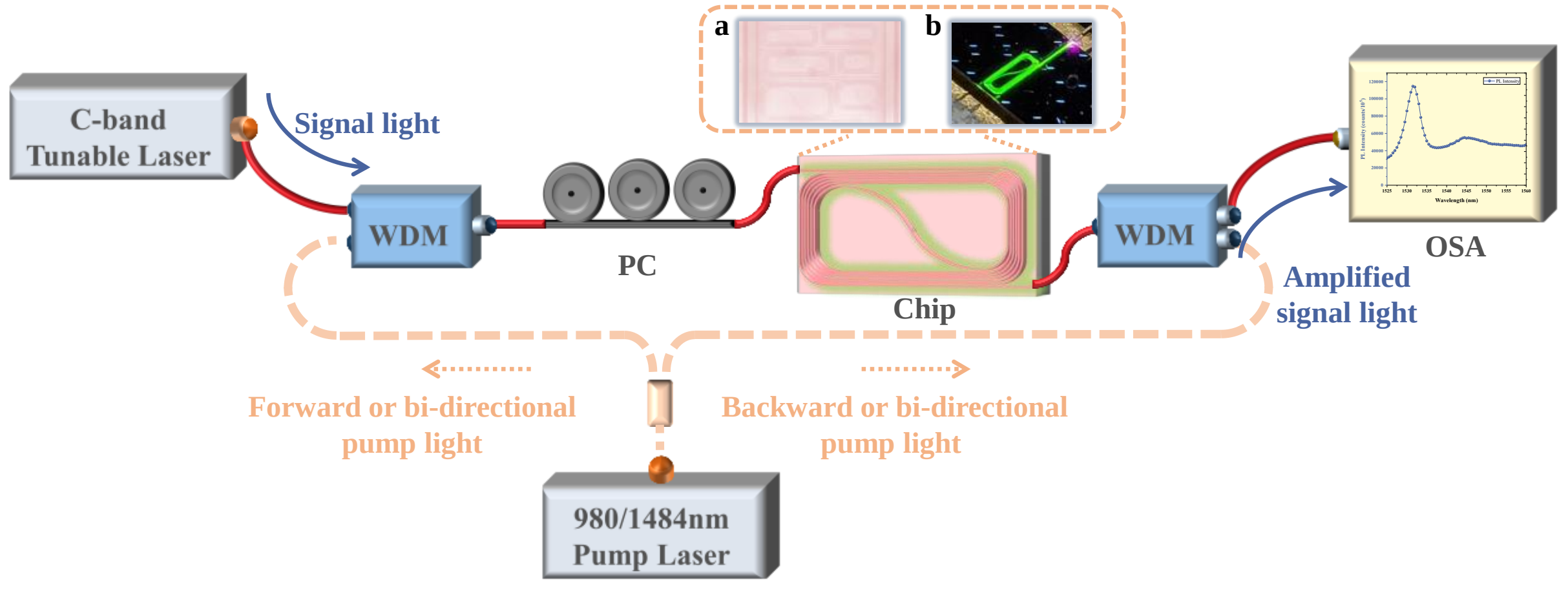

**Figure 2.** Experimental setup. (a) Micrograph of waveguide amplifiers. (b) Photo of the device with pump applied.

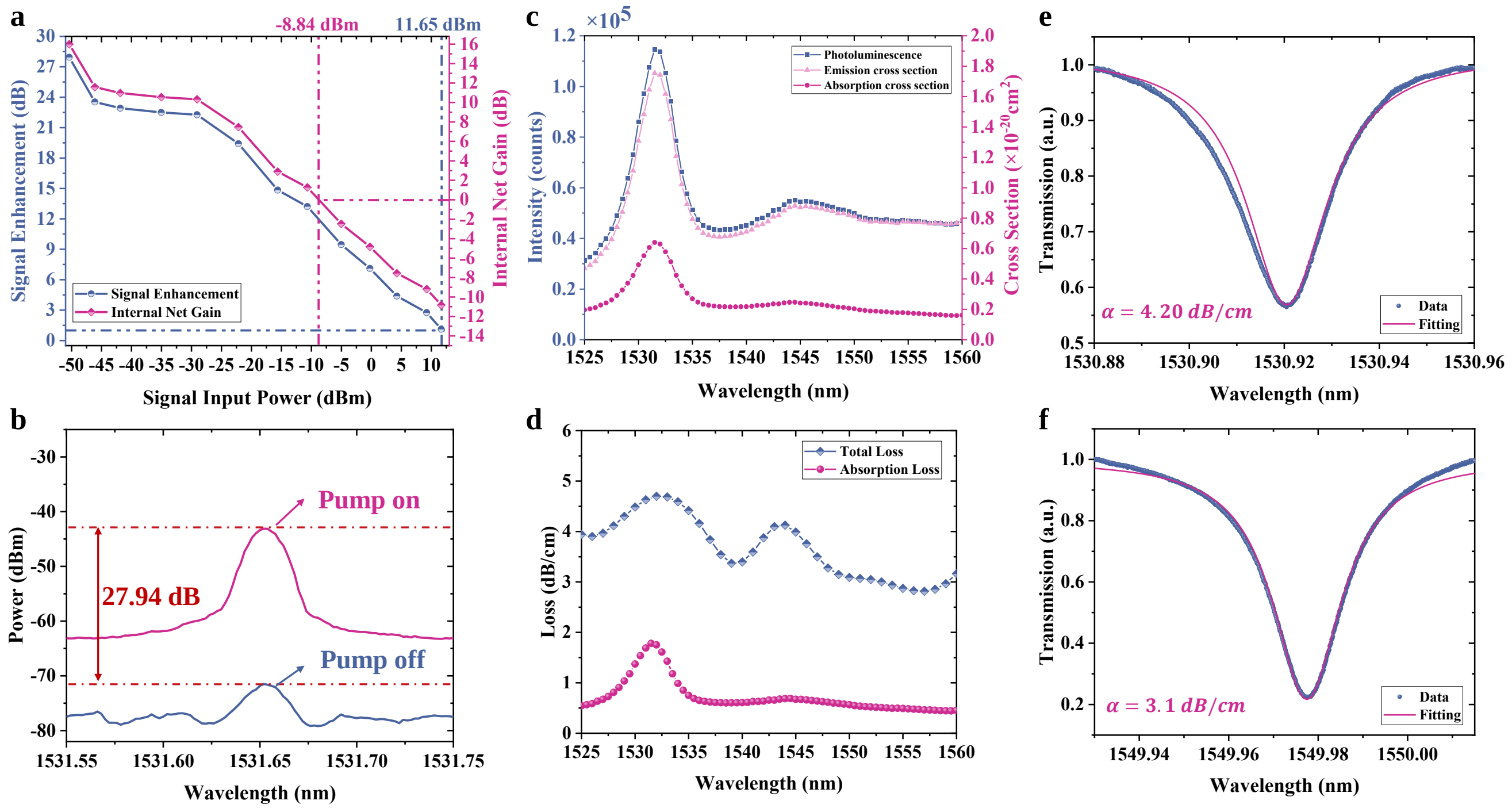

**Figure 3.** (a) Signal enhancement and internal net gain at different signal input power at 1531.6 nm. (b) Optical spectra of signals at maximal signal enhancement of 27.94 dB. (c) PL spectrum, emission cross-section, and absorption cross-section as a function of wavelength. (d) Total loss and erbium absorption loss. Transmission spectra and fitting curves of an Er:LNOI microring resonator near (e) 1531 nm and (f) 1550 nm.

The signal enhancement ($G_{se}$ in dB) of a waveguide amplifier is defined as the power contrast of the output signal when the pump power is turned on and off, given by

$$G_{se} = 10\lg(P_{on}/P_{off}), \tag{1}$$

where $P_{on}$ and $P_{off}$ are the signal power with pump power on and off, respectively. The signal enhancement was directly measured by comparing the signal power with and without pump in the OSA. At a launched (on-chip) 1484-nm pump power of 17.34 mW, a maximal signal enhancement of 27.94 dB has been achieved with -50 dBm signal input power (on-chip), as shown in Fig. 3(b). With the increase of signal input power, the signal enhancement gradually decreased and reached 1 dB at a signal input power of 11.56 dBm. The internal net gain ($G_{net}$ in dB) is defined as signal enhancement with the compensation of total loss on the chip, given by

$$G_{net} = G_{se} - \alpha_{total_loss} L, \tag{2}$$

where $\alpha_{total_loss}$ is the total loss per unit length, mainly including the erbium absorption loss and waveguide propagation loss caused by the sidewall roughness, and L is the length of waveguide amplifier. Obviously, the total loss is wavelength dependent because erbium ions have different absorption for different wavelength.

To estimate the internal net gain from the input to the output of the waveguide amplifiers, the waveguide loss should be correctly estimated. We used the following methods to estimate the erbium absorption loss and total loss.

Firstly, the erbium absorption loss was calculated from the photoluminescence (PL) spectroscopy of $Er^{3+}$, which is plotted in Fig. 3(c) with a blue dotted line. The peak near 1530 nm in the PL spectrum corresponds to the transition between the excited state and ground state. By using the Füchatbauer-Ladenburg method, the emission cross section $\sigma_e(\lambda)$ (pink line in Fig. 3(c)) can be derived from the PL spectrum through the following equation [28],

$$\frac{1}{\tau_{rad}} = 8\pi n^2 c \int \frac{\sigma_e(\lambda)}{\lambda^4} d\lambda \,, \qquad (3)$$

where $\lambda$ is the wavelength, n is the refractive index of the Er:LNOI waveguide. $\tau_{rad}$ is the radiative lifetime of $Er^{3+}$ ions, which is related to the fluorescence lifetime $\tau_{pl}$ [29] and $\sigma_e(\lambda)$ is inversely proportional to $\tau_{rad}$. For calculation, the measured $\tau_{pl}$ of the $^4I_{13/2}$ level can be assumed as purely radiative decay [30].

Combined with McCumber's theory, the absorption cross-section $\sigma_a(\lambda)$ (purple line in Fig. 3(c)) can be obtained from the emission cross section $\sigma_e(\lambda)$ as follows [31]

$$\sigma_a(\lambda) = \sigma_e(\lambda) e^{(\frac{h\nu-\varepsilon}{kT})} \,, \qquad (4)$$

where $T$ is the temperature, $h$ is the Planck's constant, $k$ is Boltzmann's constant, $\varepsilon$ is the excitation energy related to temperature, where $e^{(-\frac{\varepsilon}{kT})}=7.166\times10^{-15}$. The calculation of $\varepsilon$ can be referred to [32].The absorption loss $\alpha_a(\lambda)$ was then calculated from the following equation [28]

$$\alpha_a(\lambda) = 10\log(e)\Gamma(\lambda) N_{Er}\sigma_a(\lambda) \,, \qquad (5)$$

where $\Gamma(\lambda)$ is the overlap integral between the signal mode and erbium-doped waveguide, and the value is 0.89. $N_{Er}$ is the $Er^{3+}$ concentration in LN thin film. Through the above derivation, the erbium absorption loss from 1520 nm to 1560 nm can be obtained, as shown in pink line in Fig. 3(d). The PL spectrum has the peaks near 1530 nm and 1545 nm. Considering the gain peaks and common low propagation loss band, the signal wavelengths of 1531.6 nm and 1550 nm are paid close attention to. The erbium absorption loss at 1531.6 nm and 1550.0 nm are 1.78 dB/cm and 0.56 dB/cm.

Secondly, the total loss of the waveguide was obtained by truncation method, as shown in blue line in Fig. 3(d). It is evaluated to be 4.63 dB/cm and 3.08 dB/cm at 1531.6 nm and 1550 nm, respectively. The total loss at 1531.6 nm and 1550 nm has also been double confirmed by the resonator Q-analysis method, as shown in Fig. 3(e) and 3(f). For an Er:LNOI micro-ring resonator, the transmission spectra are fitted near 1531 nm and 1550 nm to calculate the total loss, and the calculation results are 4.20 dB/cm and 3.1 dB/cm. Noting that the measured transmission spectra were weakly distorted by the spontaneous emission excited by the frequency-sweep laser in the measurement, we made a more conservative choice of the values of total loss obtained from the truncation method, i.e., 4.63 dB/cm for 1531.6 nm and 3.08 dB/cm for 1550 nm. The difference between total loss and erbium absorption loss can be regarded as the waveguide propagation loss, which is near ~2.5 dB/cm.

With known total loss, the internal net gain can be calculated based on Eq. (2), as shown in Fig. 3(a) (pink). The maximal internal net gain is 16.0 dB at a signal power of -50 dBm. For the waveguide length of 2.58 cm, the corresponding internal net gain per unit length is 6.20 dB/cm. If we define a saturation power where the internal net gain equals to 0 dB, the saturation power is -8.84 dBm for the signal amplification at 1531.6 nm.

The signal enhancement and internal net gain at 1550 nm were also measured, as shown in Fig. 4(a). At signal input power of -50 dBm and launched pump power of 18.54 mW, the maximal signal enhancement of 10.61 dB and internal net gain of 2.66 dB were achieved. The signal enhancement reaches 1 dB at a signal power of 8.6 dBm and the saturation power of 0 dB internal net gain is -22.47 dBm. The corresponding on-chip signal output power for 1531.6 nm and 1550 nm is shown in Fig. 4(b). Since the saturation power of 0 dB internal net gain is a threshold above which the signal becomes attenuated rather than amplified, the signal output power is less than the input power after this threshold. Moreover, as the waveguide has higher absorption at 1531.6 nm than 1550 nm, the input power is attenuated more at 1531.6 nm when the input power is high, e.g., > 5dBm.

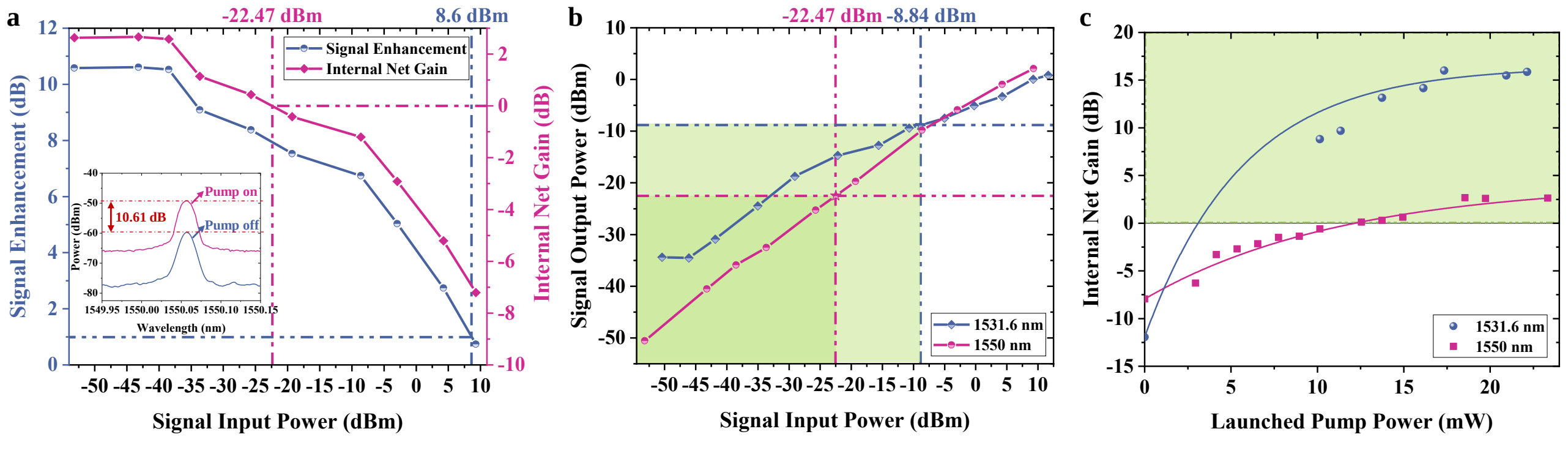

**Figure 4.** (a) Signal enhancement and internal net gain at 1550 nm. Inset: Optical spectra of signals at maximal signal enhancement. (b) Output power at 1531.6 nm and 1550 nm. (c) Internal net gain as a function of launched pump power. Green region: internal net gain greater than 0 dB.

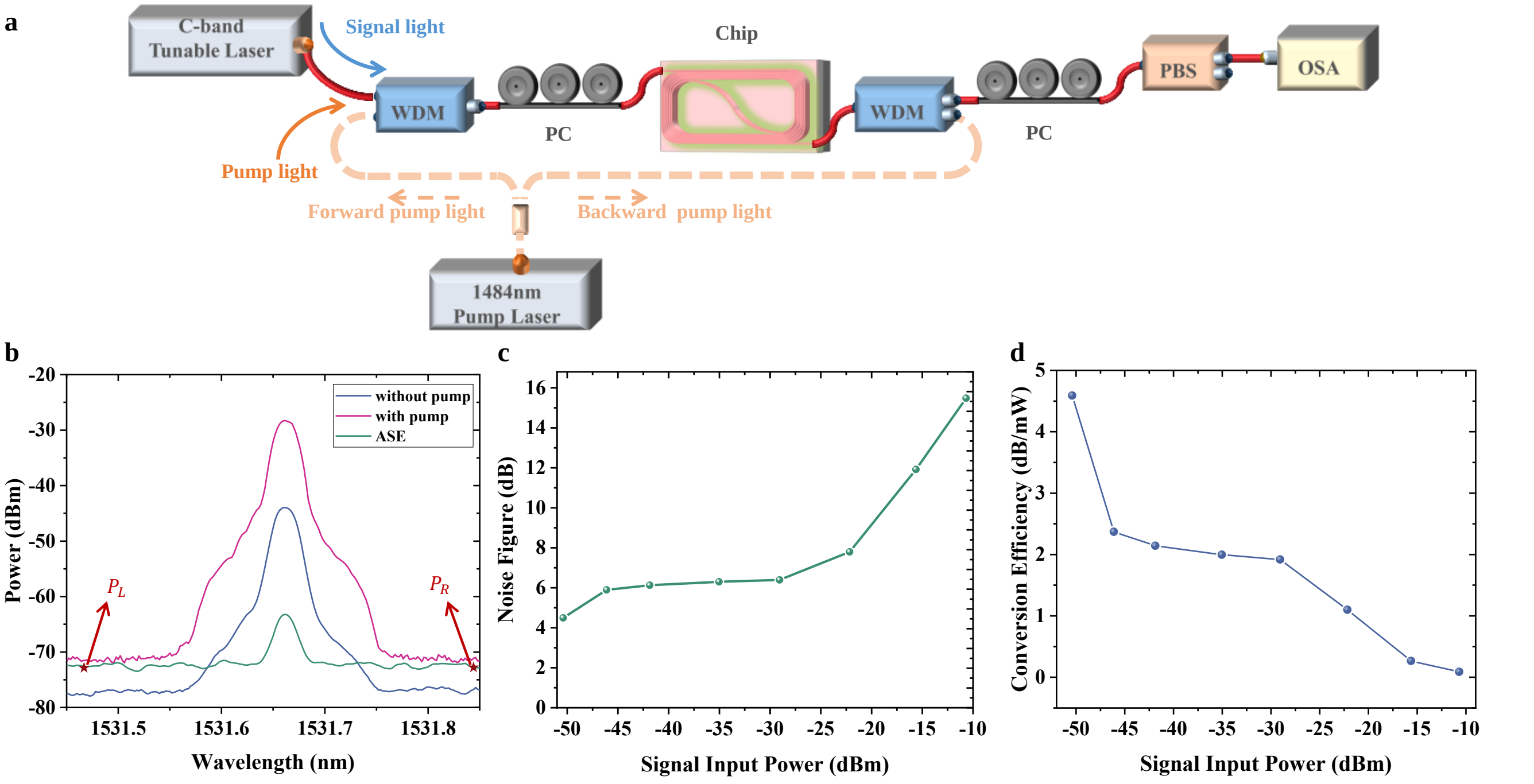

**Figure 5.** (a) Experimental setup for the measurement of noise figure. (b) Measurement of ASE power. (c) Noise figure as a function of signal input power at 1531.6 nm. (d) Measured power conversion efficiency.

The dependence on the launched (on-chip) pump power has also been investigated for the signals at 1531.6 nm and 1550 nm, as shown in Fig. 4(c). The input signal power is -50 dBm. For both wavelengths, the internal net gain gradually reaches a saturated value when the pump power is higher than 18 mW. This is because the erbium ions have been nearly all excited by the pump and the small-signal gain is therefore purely dependent on the erbium doping concentration for a given waveguide design

## B. Noise Figure and Efficiency

To evaluate the waveguide amplifier, the noise figure is a key performance index. The noise introduced by the waveguide amplifiers would strongly influence the whole system performance. The noise figure is characterized by comparing the electrical signal-to-noise ratio (SNR) between input and output, given by,

$$NF(dB) = 10\log\frac{(SNR)_{in}}{(SNR)_{out}} . \tag{6}$$

Generally, SNR degradation is dominated by shot noise and beat noise between signal and spontaneous emission. So the calculation of noise figure can be simplified to the equation [33]

$$NF = \frac{P_{ASE}}{h\nu GB} + \frac{1}{G} , \tag{7}$$

where $P_{ASE}$ is the amplified spontaneous emission (ASE) power within the measured bandwidth $B$, $h$ is the Planck's constant, $\nu$ is the optical frequency, and $G$ is the internal net gain. As the internal net gain and optical frequency have been known, only the ASE power was to be measured. The polarization extinction method was adopted to measure the noise figure.

The setup is shown in Fig. 5(a). Compared to the setup of bi-directional pumping in Fig. 2, a PC and a polarization beam splitter (PBS) were added. The ASE power were measured by adjusting the PC until the signal power in the OSA was suppressed to its minimal value. Figure 5(b) shows the measured optical spectra with minimal (green) and maximal (pink) signal power and the spectrum without pump (blue).

To estimate an accurate value of the ASE power at signal wavelength, two ASE power values of $P_L$ and $P_R$ on the left and right side of the signal wavelength with 0.2 nm interval were acquired. The ASE power $P_{ASE}$ was then calculated by

$$P_{ASE}(dB) = \frac{P_L + P_R}{2} + IL + 3 . \tag{8}$$

where IL is the insertion loss from the chip output to the OSA. 3 dB means the ASE power distributed equally in the two orthogonal polarization states. The measurement bandwidth in the OSA was 0.02 nm (~2.5 GHz).

The calculated noise figures with respect to different signal input power at 1531.6 nm are summarized in Fig. 5(c). The noise figure has a positive correlation with the signal input power. The minimal noise figure is 4.49 dB with -50 dBm small-signal power, and the flatten region from -45 dBm to -28 dBm corresponds to the NF near 6 dB. For the input signal power reaches -20 dBm, the noise figure is 8 dB and then increases rapidly with the increased signal power. This is reasonable because with the increase of signal power the gain drops significantly and the noise figure increases based on Eq. (7).

In addition, for an assessment of the effectiveness of the pump, the power conversion efficiency (PCE) is discussed at 1531.6 nm, which is defined as the ratio of internal net gain and launched pump power, $PCE = G_{net_gain}/P_{pump}$. As shown in Fig. 5(d), the maximal PCE is 4.59 dB/mW in the small-signal regime, and then decreases with the increasing signal input power. The PCE near zero corresponds to the saturation power threshold.

## 4. DISCUSSION

To establish a comprehensive understanding on the performance of Er:LNOI amplifier, various parameters have been investigated. The effect of pumping wavelength was first investigated. Different from erbium doped fiber whose index contrast is typically ~0.01 or less, the ridge Er:LNOI waveguide has a much larger index contrast of ~0.07. This leads to a mode distribution highly dependent on the wavelength. As shown in Fig. 6(a), the inset shows the optical modes at 980 nm, 1484 nm and 1530 nm, respectively. The pump at 1484 nm has a much better mode overlap with the signal at 1530 nm and thus a better amplification than the pump at 980 nm. It is compared for the signal enhancement in the C band at 1484 nm pump and 980 nm pump with the 2.58 cm-long waveguide amplifier. The signal power and pump power were fixed to 3 μW and 12.54 mW. Forward pumping scheme was adopted. It can be clearly seen that within the whole C band, 1484 nm pump exhibits advantages over the 980 nm pump including maximal signal enhancement and signal power conversion efficiency. Therefore, we chose 1484 nm as our pumping wavelength.

Then the effect of waveguide length was studied. The gain performance in the C band of four waveguides with lengths of 1.38 cm, 1.94 cm, 2.58 cm and 2.8 cm has been compared, as shown in Fig. 6(b). It can be observed that the signal enhancement increases with the increasing length from 1.38 cm to 2.58 cm and then decreases at 2.8 cm length because of the higher waveguide loss. The gain performance under different pumping schemes as a function of launched pump power has been verified in Fig. 6(c). The waveguide length is 2.8 cm and the signal power is -50dBm. With the increase of pump power, compared with other approaches, bi-directional pumping achieves the largest signal enhancement. The forward pumping provides high gain for a small pump power (e.g., 10 mW) but becomes less effective when the pump power is higher. This is because the pump power has opposite distribution (high at beginning of waveguide and low in the end) compared to the signal power (low at beginning and high in the end). The backward pumping solves this problem because both pump and signal have low power at beginning of waveguide and high power in the end. But it is known that backward pumping typically introduces higher noise [34]. Therefore, bi-directional pumping was chosen in our experiment to find a balance between the gain and noise performance [34].

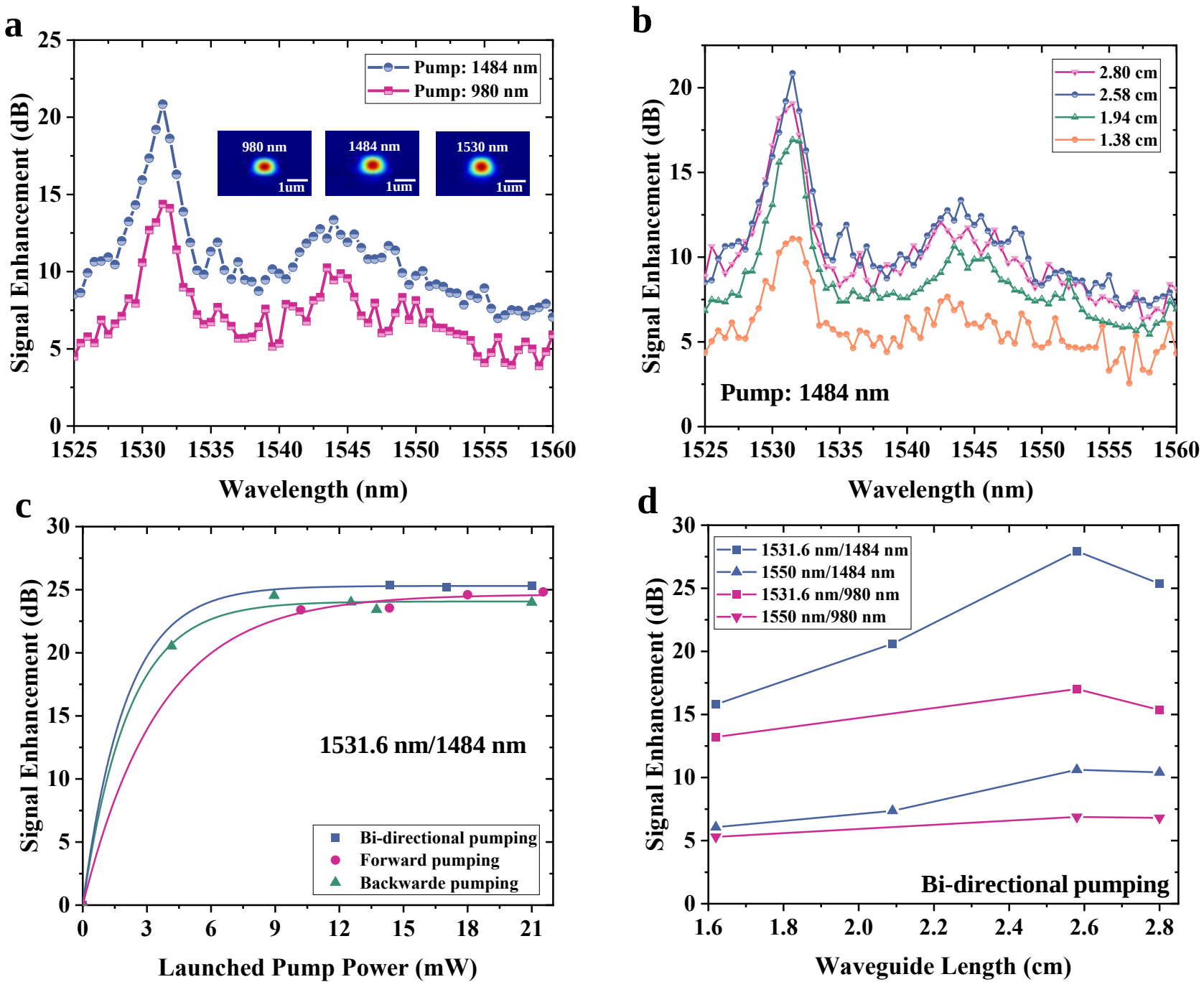

**Figure 6.** (a) Comparison of signal enhancement between 980 nm and 1484 nm pump light. Inset: Mode profiles in the Er:LNOI waveguide at 980 nm, 1484 nm, and 1530 nm. (b) Signal enhancement as a function of signal wavelength with 1484 nm pump light at different waveguide lengths. (c) Signal enhancement as a function of launched pump power by three pumping approaches at 1531.6 nm with 1484 nm pump. (d) Signal enhancement with respect to different waveguide lengths under bi-directional pumping with 980 nm and 1484 nm pump light.

**Table 1. Parameter comparison of recent works on Er:LNOI amplifiers.**

| Signal wavelength | Length | Pump wavelength | Pump power | Signal enhancement | Internal net gain | Ref |
|---|---|---|---|---|---|---|
| 1531.6 nm | 5 mm | 980 nm | 21 mW | 6.2 dB | 5.2 dB | 23 |
| 1530 nm | 5.3 mm | 980 nm | 10.78 mW | 11.94 dB | 8.3 dB | 24 |
| 1531.5 nm | 5 mm | 974 nm | 61 mW | 15.415 dB | 15 dB | 25 |
| 1530 nm | 3.6 cm | 980 nm | 40 mW | 18.756 dB | 18 dB | 26 |
| 1531.6 nm | 2.8 cm | 980 nm | 45 mW | 22.2 dB | 8.45 dB | 35 |
| 1531.6 nm | 2.58 cm | 1484 nm | 17.31 mW | 27.94 dB | 16 dB | This work |

Figure 9(d) shows the maximal signal enhancement under the combination of different signal wavelength, pump wavelength and waveguide length. The signal input power was fixed to -50 dBm and bi-directional pumping was adopted. It can be found that signal at 1531.6 nm with 1484 nm pump typically has better gain performance than 1550 nm signal with 980 nm pump. The maximal signal enhancement of 27.94 dB appears under the condition of -50 dBm signal power at 1531.6 nm, 17.31 mW pump power at 1484 nm and 2.58-cm-long waveguide.

Finally, Er:LNOI is a promising platform thanks to the versatile property combination of LN and erbium. Er:LNOI amplifiers have shown superior performance recently due to the high doping concentration, good mode confinement and low-loss waveguide. In Table 1, we have summarized the state-of-art works of Er:LNOI amplifiers in terms of signal enhancement and internal net gain. Benefited from optimized waveguide design, pumping wavelength and pumping scheme, our work has achieved a maximal signal enhancement of ~28 dB and internal net gain of ~16 dB with only ~17 mW pump power and 2.58 cm waveguide length, indicating a highly efficient and compact waveguide amplifier design. It is believed that, by further optimizing the waveguide fabrication technology and reducing the waveguide propagation loss, these specifications can be further improved.

## 5. CONCLUSION

In conclusion, we have demonstrated a high-efficiency erbium-doped waveguide amplifier on the LNOI platform. The comprehensive understanding of the device has been built up, including the emission/absorption cross section, output saturation, effect of pump wavelength and pumping scheme, effect of waveguide length, and noise performance. With the full knowledge of this device, a maximal signal enhancement of 27.94 dB and internal net gain of 15.99 dB have been achieved at 1531.6 nm on a 2.58 cm-long Er:LNOI waveguide and 17.31 mW pump power. This indicates a highly efficient and compact design of an integrated waveguide amplifier. Moreover, the noise figure is only 4.49 dB under a signal power of -50 dBm, indicating the low-noise operation under small-signal regime. The saturation power is -8.84 dBm where internal net gain equals to 0 dB. Our work has demonstrated that Er:LNOI can be a powerful platform to achieve integrated waveguide amplifiers with superior performance. Together with the abundant photonic and optoelectronic properties of LNOI, a fully integrated photonic system on LNOI can be envisaged.

## AUTHOR INFORMATION

### Authors

**Minglu Cai, Kan Wu, Junmin Xiang, Zeyu Xiao, Tieyin Li, Chao Li, and Jianping Chen**- State Key Laboratory of Advanced Optical Communication Systems and Networks, Department of Electronic Engineering, Shanghai Jiao Tong University, Shanghai 200240, China

Corresponding email: *kanwu@sjtu.edu.cn*

### Notes

The authors declare no competing financial interest.

### Funding Sources

National Natural Science Foundation of China (NSFC) (No. 61922056, 61875122).

## ACKNOWLEDGMENT

We acknowledge The instrumental analysis center of Shanghai Jiao Tong University for XPS and SIMS analysis, and the center for advanced electronic materials and devices (AEMD) of Shanghai Jiao Tong University for fabrication support.